\documentclass[]{svproc}
\usepackage{url}

\usepackage[latin9]{inputenc}
\usepackage{amsmath}
\usepackage{amssymb}
\usepackage{graphicx}
\usepackage{bbm}

\usepackage[usenames,dvipsnames,svgnames]{xcolor}
\definecolor{myurlcolor}{rgb}{0,0,0.7}
\definecolor{myrefcolor}{rgb}{0.8,0,0}
\usepackage[unicode=true,pdfusetitle, bookmarks=true,bookmarksnumbered=false,bookmarksopen=false, breaklinks=false,pdfborder={0 0 0},backref=false,colorlinks=true, linkcolor=myrefcolor,citecolor=myurlcolor,urlcolor=myurlcolor]{hyperref}

\makeatletter
\@ifundefined{textcolor}{}
{%
 \definecolor{BLACK}{gray}{0}
 \definecolor{WHITE}{gray}{1}
 \definecolor{RED}{rgb}{1,0,0}
 \definecolor{GREEN}{rgb}{0,1,0}
 \definecolor{BLUE}{rgb}{0,0,1}
 \definecolor{CYAN}{cmyk}{1,0,0,0}
 \definecolor{MAGENTA}{cmyk}{0,1,0,0}
 \definecolor{YELLOW}{cmyk}{0,0,1,0}
}

\usepackage{txfonts}

\newcommand{\ket}[1]{\left| {#1} \right\rangle}
\newcommand{\bra}[1]{\left\langle {#1}\right|}

\newcommand{\eins}{\mathbbm{1}}
\newcommand{\id}{\mathcal{I}}

\renewcommand{\t}[1]{\textrm{#1}}
\renewcommand{\sp}{{\sigma}}

\newcommand{\Cang}{{\vartheta}}
\newcommand{\tr}{{\mathrm{Tr}}}

\newcommand\dynmat{\mathsf{D}}

\renewcommand{\vec}[1]{\boldsymbol{#1}}
\newcommand{\dd}[1]{\frac{\mathrm{d}}{\mathrm{d}#1}}
\newcommand{\wS}{{\omega_0}}
\newcommand{\sz}{{\sigma_z}}
\newcommand{\sy}{{\sigma_y}}
\newcommand{\sx}{{\sigma_x}}
\newcommand{\sbar}{{\bar{\sigma}}}
\newcommand{\e}{{\mathrm{exp}}}

\newcommand{\eqnref}[1]{Equation~(\ref{#1})}

\newcommand{\figref}[1]{Figure~\ref{#1}}

\makeatother

\begin{document}
\mainmatter              
\title{Non-monotonic population and coherence evolution in Markovian open-system dynamics}
\titlerunning{Non-monotonic population and coherence evolution in Markovian dynamics}

\author{J.~F.~Haase\inst{1}, A. Smirne\inst{1} and S.~F.~Huelga\inst{1}}

\institute{Institut f{\"u}r Theoretische Physik and IQST, Albert-Einstein-Allee
11, Universit{\"a}t Ulm, D-89069 Ulm, Germany}

\authorrunning{J.~F.~Haase et al.}

\maketitle

\begin{abstract}
We consider a simple microscopic model where the open-system dynamics of a qubit,
{despite being Markovian,
shows features which are typically associated to the presence of memory effects.
Namely,} a non monotonic behavior both in the population and in the coherence evolution
{arises due to the presence of non-secular contributions, which
break the phase covariance of the Lindbladian (semigroup) dynamics.
We also show by an explicit construction how such a non-monotonic
behaviour can be reproduced by a phase covariant evolution, but
only at the price of inserting some state-dependent memory effects.}

\end{abstract}

\section{Introduction}

Non-Markovianity in the evolution of open quantum systems \cite{Breuer2002,Rivas2012}
{has been attracting a renewed interest in the last years, both from a fundamental point of view \cite{Rivas2014,Breuer2016,deVega2017,Li2017} and
also in the light of its possible role}
in different areas of upcoming quantum technologies. Memory effects may yield to improved performance in quantum teleportation \cite{Laine2014,Torre2018}, quantum key distribution \cite{Vasile2011} or superdense coding \cite{Liu2016},
{as well as enhanced capacity of quantum channels} \cite{Bylicka2014} and assist
preparation of entangled states in the steady state \cite{Plenio1999,Plenio2008}.
Furthermore, temporal correlations in the environment are a necessity for noise-refocusing protocols frequently employed in quantum sensing \cite{Cywinski2008,Lange2011,Degen2017}.
Precision bounds in quantum metrology have also been shown to depend on the nature of the environmental noise \cite{Chin2012,Benjamin2011} and proven to be insensitive to memory effects in the asymptotic limit \cite{Smirne2016,Haase2018Metro}.
In the quest of revealing and controlling
{the degree of non-Markovianity} in different experimental architectures for quantum computation, quantum information and quantum simulation tasks, experiments have been performed in optical setups \cite{Liu2011,Cialdi2011,Chiuri2012,Liu2013,Jin2015,Cialdi2017}, ion traps \cite{Wittemer2018} and nitrogen-vacancy centers in diamond \cite{Haase2018,Wang2018,Peng2018}.
{Moreover, the characterization of complex open-system dynamics, involving
memory effects due to
the interaction with structured environments,
has led to the development of novel analytical and numerical tools \cite{Tanimura1989,Diosi1998,Prior2010,Ciccarello2013,Diosi2014,Batalhao2014,Vacchini2016,Gasbarri2017,Tamascelli2018,Campbell2018}
beyond the standard perturbative techniques \cite{vanKampen1974,Shibata1977,Chaturvedi1979,Breuer2001}.}

Whether the evolution of an open quantum system {is Markovian or not 
generally depends on the chosen definition \cite{Rivas2014,Breuer2016,Li2017}.
Here, we will focus in particular on} the criteria of, respectively, CP-divisibility \cite{Rivas2010} (usually refered to as the RHP definition) and monotonicity of the trace distance \cite{Breuer2009} (the so-called BLP definition).
While {both these definitions have a clear physical meaning and can be formulated straightforwardly from
a mathematical point of view, their} experimental verification can be a challenging task, since it either requires process tomography (RHP) or involves a minimization over different input states (BLP).
{Now,} since Markovianity was historically associated
with the exponential evolutions of populations and coherences of the open quantum system states,
one can be tempted to ascribe non-monotonic behaviours
of populations and coherences to deviations from Markovianity
\cite{Madsen2011,Duan2017,Passos2018}, identifying them as witnesses of non-Markovianity.

In this work we present an exactly solvable model, {obtained
via a weak-coupling microscopic derivation \cite{Breuer2002}}, which acts as a counterexample to the previous conclusion. 
In fact, the dynamics of our model shows non-monotonic behaviors in populations and coherences, while
being generated by a Lindblad master equation with constant coefficients,
{i.e., being (time-homogeneous) Markovian according to both RHP and BLP definitions.
We show that this traces back to the violation of a symmetry of the dynamics -- the so called phase covariance \cite{Holevo1993,Holevo1996,Vacchini2010}. In a complementary way,
we also build up a phase covariant dynamics reproducing the non-monotonic coherence evolution,
but this time due to the presence of non-Markovian features which depend on the choice of the initial state.
Our results show how the standard intuition about (non-)Markovianity
can be recovered only under strict constraints, while
the precise verification of such property calls unavoidably for the comparison of
the evolutions obtained from different initial conditions.}

\section{Tools}
\subsection{Evolution of Open Quantum Systems}
\label{sec:OQS}
We consider the finite dimensional quantum system of interest $S$ to be coupled with an environment $E$.
The state $\rho_S$ is given by the partial trace over the environmental degrees of freedom applied to the total state of $S$ and $E$, $\rho_{SE}$, i.e., $\rho_S = \t{tr}_E \{\rho_{SE}\}$.
Fixing the initial time $t_0$, the state at time $t$ {is given by the} dynamical map $\Lambda_{t,t_0}$
according to
\begin{eqnarray}
\rho_S(t) &=& \Lambda_{t,t_0}[\rho_{S}(t_0)].
\label{eq:map}
\end{eqnarray}
The map needs to fulfill at least the properties of
\begin{enumerate}
\item positivity (P), i.e. for any positive operator $O$ it holds $O\geq 0 \Rightarrow \Lambda[O] \geq 0$, and
\item trace preservation (TP), i.e. $\t{tr}\left\lbrace \Lambda[O]\right\rbrace = \t{tr}\{O\}$.
\end{enumerate}
These properties ensure that $\rho_S(t)$ is a physical state,
{whenever the initial $\rho_S(t_0)$ is so.
However in case the open system is
initially} entangled with a further system, e.g. an ancilla $A$, the stronger property of complete positivity (CP) is required,
in order to {guarantee a meaningful evolution of} the joint system {$S+A$}.
In particular, {CP means that} $\tilde{O}\geq 0 \Rightarrow \left(\Lambda\otimes \id_d\right)[\tilde{O}] \geq 0$, where $\tilde{O}$ is a positive operator on the extended Hilbert space {associated with $S+A$ and the ancilla}
Hilbert space has the arbitrary dimension $d$ and the associated identity map $\id_d$.

In case the dynamics of the reduced system is described via a time-local master equation (ME)
$\dd{t} \rho_S(t) = \mathcal{L}_t [\rho_s(t)]$, \eqnref{eq:map} is the corresponding
solution with the initial condition $\rho_S(t_0)$.
With the standard form of the generator $\mathcal{L}_t$ {\cite{Gorini1976,Hall2014}, the most general}
ME takes the form
\begin{eqnarray}
&&\dd{t}\rho_S(t) = \mathcal{L}_t [\rho_S(t)]= -i[H(t),\rho_S(t)] \notag \\ &&+ \sum_k \gamma_k(t) \left[ V_k(t)\rho_S(t)V_k^\dagger(t)-\frac{1}{2}\lbrace V_k^\dagger(t)V_k(t),\rho_S(t)\rbrace\right],
\label{eq:ME}
\end{eqnarray}
where the $V_k(t)$ are time dependent, not necessarily Hermitian operators,
$H(t)=H^{\dagger}(t)$ is a time-dependent Hamiltonian and $\gamma_k(t)$ are time dependent rates.
In the special case where {operators and rates are instead
constant and the latter are positive}, the ME reduces to the
well-known ME in Lindblad form \cite{Gorini1976,Lindblad1976}.

Equivalently, one can consider a ME for the dynamical map itself, so that
\begin{eqnarray}
\dd{t} \Lambda_{t,t_0} = \mathcal{L}_t \Lambda_{t,t_0},
\label{eq:MeMap}
\end{eqnarray}
where the initial condition is given by $\Lambda_{t_0,t_0}=\eins$ and a formal solution can be written via a time-ordered exponential $\Lambda_{t,t_0} = \mathcal{T}_{\leftarrow} \e\left(\int_{t_0}^t \mathcal{L}_\tau\mathrm{d}\tau\right)$.

\subsection{Non-Markovianity}\label{sec:nm}
The Markovianity definition introduced by Rivas, Huelga and Plenio (RHP) is based on the CP-divisibility of dynamical maps \cite{Rivas2010}.
{The latter property means that the dynamical maps in \eqnref{eq:map} are
not only CPTP, but can always be decomposed as}
\begin{eqnarray}
\Lambda_{t_2,t_0} = \Phi_{t_2,t_1} \Lambda_{t_1,t_0} \quad \forall t_2 \geq t_1 \geq t_0,
\label{eq:mapComposition}
\end{eqnarray}
{where $\Phi_{t_2,t_1}$ is in turn a CPTP map, which can be identified with the propagator
of the dynamics under proper conditions \cite{Rivas2010}}.
In particular, {it can be shown \cite{Laine2010,Rivas2012}
that under some regularity conditions (see also \cite{Chruscinski2018})
CP-divisibility is equivalent to the presence of a}
generator $\mathcal{L}_t$ such that $\gamma_k(t)\geq 0$ at any arbitrary point in time.

{An even stronger condition for a Markovian evolution corresponds to the case of CP semigroup dynamics,
for which both the dynamical maps and  {the propagators depend only} on the difference of their time arguments,
$\Lambda_{t,t_0}=\Lambda_{t-t_0,0}\equiv \Lambda_{t-t_0}$
and $\Phi_{t_2,t_1}=\Lambda_{t_2-t_1}$, so that
the composition in \eqnref{eq:mapComposition} simplifies to}
\begin{eqnarray}
\Lambda_{t+s} = \Lambda_{t} \Lambda_{s} \quad \forall t,s\geq 0,
\label{eq:mapSemigroup}
\end{eqnarray}
where of course each of these maps is CPTP.
In addition, semigroup dynamics
are characterized by a time-independent generator $\mathcal{L}_t \equiv \mathcal{L}$ with positive rates, i.e. the mentioned Lindblad generator.
{They are thus identified with the time-homogeneous Markovian evolutions,
while CP-divisible, but not Lindbladian dynamics are the time-inhomogeneous Markovian ones;
dynamics which are not CP-divisible are non-Markovian in this framework}.

Note that the CP of $\Phi_{t_2,t_1}$ can be checked via the positivity of its Choi state
\begin{eqnarray}
\rho_C = \left(\Phi_{t_2,t_1}\otimes \id_{d_S}\right)[\ket{\psi}\bra{\psi}],
\end{eqnarray}
where $d_S$ is the dimension of the Hilbert space of $S$ and $\ket{\psi}=\frac{1}{\sqrt{d_S}}\sum_{n=1}^{d_S} \ket{n n}_{SA}$ is a maximally entangled state on $S$ and the extending Ancilla space $A$.

{A different notion of Markovianity has been introduced by Breuer, Laine and Piilo (BLP) \cite{Breuer2009}
and relies on the evolution of the trace distance. Given two states $\rho^1$
and $\rho^2$, their trace distance is defined as
\begin{equation}\label{eq:td}
d = \frac{1}{2}\| \rho^1-\rho^2\|_1 = \frac{1}{2}\sum_i |x_i|,
\end{equation}
where $\| \cdot\|_1$ denotes the trace norm and $x_i$ the eigenvalues
of the traceless operator $\rho^1-\rho^2$,
and it quantifies their distinguishability.
A non-monotonic time evolution of the trace distance $d(t)$
between two open-system states $\rho^1_S(t)$ and $\rho^2_S(t)$
evolved from two different initial states detects
a back-flow of information to the open system, resulting in an increased
amount of information about the initial condition.
In particular, any CP-divisible dynamics implies a monotonic non-increasing evolution
of $d(t)$
for any couple of initial conditions and intervals of time, but the converse implication
does not hold \cite{Laine2010,Rivas2012}.
Hence, assuming CP-divisibility as the definition of Markovianity,
we can consider a non-monotonic behavior of the trace distance
as a witness for non-Markovianity. As will become clear from the following discussion,
a crucial point for establishing the link between trace distance and Markovianity is that one is comparing
a property of the resulting evolution
for couples of different initial conditions rather than focusing on a single evolution, since the latter would not allow
to witness non-Markovianity without any further information
about the properties of the dynamics.}

\subsection{Matrix Representation of Qubit Maps}
Any qubit map possesses a convenient matrix representation that allows for the intuitive elucidation of geometrical features in the dynamics. Fixing the operator basis $\lbrace \sigma_k \rbrace_{k=0}^{3}$ where $\sigma_0=\eins$ and $\sigma_k,\, k=1,2,3$ are the Pauli matrices, we make use of the scalar product of operators $\langle \sigma,\tau \rangle = \tr \lbrace \sigma^\dagger \tau \rbrace/2$ and write the action of the map as
\cite{King2001,Andersson2007,Smirne2010}
\begin{eqnarray}
\Lambda[\tau] = \sum_{k,l=0}^3 \dynmat^\Lambda_{kl} \langle \sigma_l, \tau \rangle \sigma_k,
\end{eqnarray}
where
\begin{equation}
\dynmat^\Lambda_{kl} = \langle \sigma_k, \Lambda[\sigma_l]\rangle
\end{equation}
is the $4 \times 4$ matrix representation of the map.
{Due to the particular choice of the operator basis, the matrix} takes the general form
\begin{eqnarray}
\dynmat^\Lambda = \left(\begin{array}{cc}
1 & \vec{0}^\mathrm{T} \\
\vec{v} & V
\end{array} \right),
\label{eq:DynMat_MatrixRep}
\end{eqnarray}
with a real three element column vector $\vec{v}$, a three element zero row vector $\vec{0}^\mathrm{T}$ and a real
$3 \times 3$ matrix $V$
{(as a consequence of trace and Hermiticity preservation)}.
This representation provides us with a clear geometrical picture of the evolution of the qubit's Bloch vector $\vec{S}$
{associated to a generic state $\rho$
via the relation $\rho=\left[\eins +  \vec{S} \vec{\sigma}\right]/2$},
where $\vec{\sigma}$ is the standard vector of Pauli matrices. Under the action of the map, the state evolves as
\begin{eqnarray}
\Lambda[\rho]=\frac{1}{2}\left[\eins + ( \vec{v} + V \vec{S}) \vec{\sigma}\right],
\label{eq:BochStateTransformed}
\end{eqnarray}
{or, in other terms, one has the affine transformation $\vec{S} \mapsto  \vec{v} + V \vec{S}$}:
$\vec{v}$ causes translations of the Bloch vector, while $V$ introduces rotations, contractions and reflections of $\vec{S}$.
For example, the qubit's phase evolution corresponds to a rotation of $\vec{S}$ around the $z$ axis of the Bloch sphere.

\subsection{Phase Covariance of Open System Evolutions}
Let us briefly introduce the concept of phase covariance \cite{Holevo1993,Holevo1996,Vacchini2010},
{which is sometimes also indicated as time-translation symmetry
in the literature \cite{Marvian2016,Lostaglio2017}}.
Mathematically, it can be characterized by a commutation relation of dynamical maps.
For the sake of simplicity we restrict here to qubit maps.
The dynamics fixed by $\Lambda_{t_1,t_0}$ is said to be phase covariant (PC), if
$\forall \phi \in \mathbbm{R}$,
\begin{eqnarray}
\mathcal{U}_\phi \circ \Lambda_{t_1,t_0} =  \Lambda_{t_1,t_0} \circ \mathcal{U}_\phi \;\; \t{with}\;\; \mathcal{U}_\phi[\bullet] = \mathrm{e}^{-i\phi \sz} \bullet \mathrm{e}^{i\phi \sz}.
\end{eqnarray}
Crucially, this relation has a clear physical meaning, as it holds true when
the secular approximation on the generator of the ME is valid \cite{Haase2018Metro},
{which implies that} the evolution of populations and coherence are decoupled.
In particular, any PC qubit ME can be written as \cite{Smirne2016}
\begin{eqnarray}
\dd{t}\rho_S(t)&=& -i\left[\frac{\omega_0+h(t)}{2} \sp_z ,\rho_S(t)\right] \notag \\
&& + \gamma_{+}(t)\left(\sp_+ \rho_S(t) \sp_- -\frac{1}{2}\left\lbrace\sp_- \sp_+,\rho_S(t)\right\rbrace\right) \notag \\
&& + \gamma_{-}(t)\left(\sp_- \rho_S(t) \sp_+ -\frac{1}{2}\left\lbrace\sp_+ \sp_-,\rho_S(t)\right\rbrace\right) \notag \\
&& + \gamma_{z}(t)\left(\sp_z\rho_S(t) \sp_z - \rho_S(t)\right),
\label{eq:SecMasterEquation}
\end{eqnarray}
with the absorption and emission rates $\gamma_+(t)$ and $\gamma_-(t)$, the dephasing rate $\gamma_z(t)$,
{and for future convenience we have separated
a free Hamiltonian contribution fixed by $\omega_0$ and} the Lamb shift $h(t)$.
Conversely, any generator which can not be written in the form above, is non phase covariant (NPC).

\begin{figure}[t!]
 \includegraphics[width=1\columnwidth]{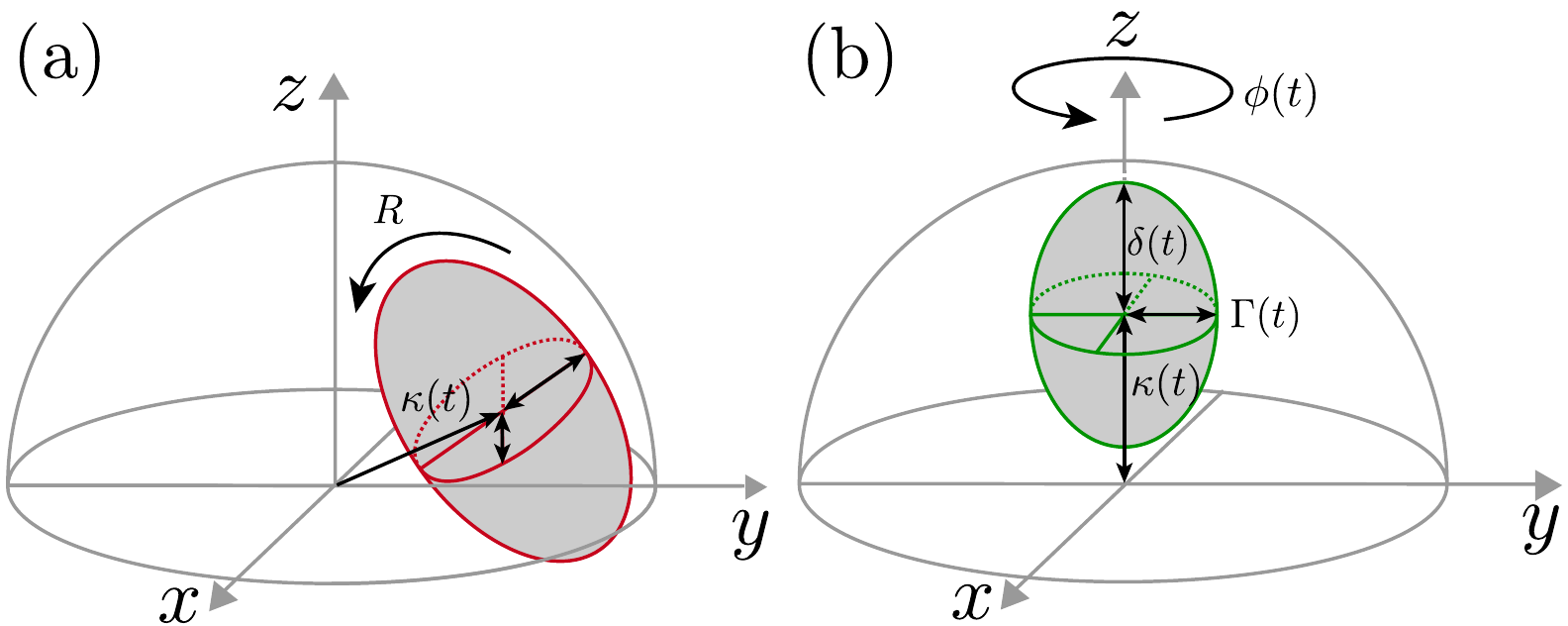}
\caption{%
\textbf{Transformation of the Bloch Sphere under PC and NPC dynamics.} The grey areas mark the volume of possible states after the application of a dynamical map. General NPC dynamics are depicted in panel a). They include arbitrary translations, rotations and contractions and further include reflections. Contrary, PC dynamics as illustrated in panel b) are restricted to transformations preserving the rotational symmetry of the volume and only include translations and contractions along $z$, equal contractions along $x$ and $y$, rotations around $z$ {and reflections about the $x-y$ plane}.
}
\label{fig:FigSpheres}
\end{figure}

The map generating the evolution dictated by the ME in eq. \eqref{eq:SecMasterEquation} can be written explicitly
{in the matrix form defined in \eqnref{eq:DynMat_MatrixRep}} as \cite{Smirne2016}
\begin{eqnarray}
\dynmat^\Lambda_{t}=\left(\begin{array}{cccc}
1 & 0 & 0 & 0 \\
0 & e^{\Gamma(t)} \cos \phi(t) & -e^{\Gamma(t)} \sin \phi(t) & 0 \\
0 & e^{\Gamma(t)} \sin \phi(t) & e^{\Gamma(t)} \cos \phi(t) & 0 \\
\kappa(t) & 0 & 0 & e^{\delta(t)}
\end{array} \right),
\label{eq:PC}
\end{eqnarray}
where we omitted the initial time $t_0$. The defined quantities are given by
\begin{eqnarray}
\phi(t) &=& \int_{t_0}^t h(\tau) \, \mathrm{d}\tau + \wS (t-t_0) \notag \\
\Gamma(t) &=& -\frac{1}{2} \int_{t_0}^t \left[\gamma_+(\tau) + \gamma_-(\tau) +4 \gamma_z(\tau)\right] \,\mathrm{d}\tau  \notag \\
\kappa(t) &=& \int_{t_0}^t \left [ \gamma_+(\tau) - \gamma_-(\tau) \right]  \notag \\
&& \times \exp \left(-\int_{\tau}^t  \left [ \gamma_+(s) + \gamma_-(s) \right] \mathrm{d}s \right) \, \mathrm{d}\tau \notag
\\
\delta(t) &=& -\int_{t_0}^t  \left [ \gamma_+(\tau) + \gamma_-(\tau) \right] \mathrm{d}\tau.
\label{eq:generalSolsPC}
\end{eqnarray}
From the matrix representation we can deduce the geometrical transformations that
PC qubit maps perform on the Bloch vector, as illustrated pictorically in \figref{fig:FigSpheres}.
These consist of translations (contractions) along the $z$ axis given by $\kappa(t)$ ($e^{\delta(t)}$), isotropic contractions along the $x$ and $y$ direction fixed by $e^{\Gamma(t)}$ and lastly rotations around the $z$ axis by an angle $\phi(t)$ which assembles the phase evolution.
We remark that all those transformations preserve the Bloch sphere's rotational symmetry around the $z$ axis, which clearly separates them from the NPC maps.
The latter {in fact include} non-isotropic contractions in all three directions, translations along an arbitrary vector $\vec{v}$ [see \eqnref{eq:DynMat_MatrixRep}] {not parallel to the $z$ axis, as well as} rotations around arbitrary directions.

\section{Results}
\subsection{The Microscopic Model}
\label{sec:Model}
In the following, we will work out a {simple example showing
non-monotonic features in the coherence and population} dynamics,
while at the same time being described by a Lindblad master equation with constant positive coefficients.

We consider a two-level system interacting with an infinite number of independent harmonic oscillators, i.e. the spin-boson model \cite{Leggett1987}, which is widely used to describe noise processes in open systems. The Hamiltonian is given as the sum of the free terms $H_S$ and $H_E$ of system and environment and the interaction $H_I$, such that
\begin{eqnarray}
H &=& H_0+H_E+H_I = \frac{\wS}{2}\sz + \sum_k{\omega_k}a^\dagger_k a_k \notag \\&+& \left( \frac{\cos \Cang}{2} \sx + \frac{\sin \Cang}{2} \sz \right) \otimes \sum_k \left( g_k a_k + g_k^* a_k^\dagger \right).
\label{eq:Hamiltonian}
\end{eqnarray}
Following \cite{Breuer2001,Breuer2002} we derive a time-convolutionless master equation, as we sketch in the following (a detailed derivation can be found in \cite{Haase2018Metro}).

Under the assumption that the coupling between the system and environment is weak, it is sufficient to consider the time-convolutionsless expansion to second order. In the interaction picture with respect to the free Hamiltonian, $H_0+H_E$, it is  given by (denoting interaction picture quantities with a tilde $\tilde{\bullet}$)
\begin{eqnarray}
\dd{t}\tilde{\rho}_S(t) = -\int_0^t \tr_E \left \lbrace [\tilde{H}_I(t),[\tilde{H}_I(\tau),\tilde{\rho}_S(t) \otimes \rho_E ]]\right \rbrace \mathrm{d}\tau.
\label{eq:TCLME}
\end{eqnarray}
At second order, the reduced dynamics is solely governed by the two time correlation functions $C(t,\tau)=\tr_E[B(t)B(\tau)\rho_E]$ of the environmental coupling operator in the interaction picture,
\begin{eqnarray}
B(t) =  \sum_k \left( g_k a_k e^{-i\omega_k t} + g_k^* a_k^\dagger e^{i\omega_k t} \right).
\end{eqnarray}
Assuming that $\rho_E$ is initially thermal at inverse temperature $\beta = 1/k_BT$, the two point correlation functions are determined by the difference of the time points, i.e., $C(t,\tau)\equiv C(t-\tau)$, which ultimately leads to
\begin{eqnarray}
C(t) = \int_{-\infty}^\infty e^{i \omega t} N(\omega) [J(\omega)\Theta(\omega) - J(-\omega) \Theta(-\omega)] \mathrm{d}\omega,
\end{eqnarray}
{where
\begin{equation}
N(\omega)=\frac{1}{e^{\beta \omega}-1}
\end{equation}
is the average number of excitations in a mode of frequency $\omega$
and we assumed the continuous limit in the spectral density $J(\omega)=\sum_k g_k^2 \delta(\omega-\omega_k)$.}
Furthermore, the Heaviside stepfunction $\Theta(\omega)$ ensures that the argument of $J(\omega)$ is always positive.

If the temperature of the environment is chosen to be sufficiently high, we have $N(\omega)\approx N(-\omega)$, which ensures that the corresponding map is unital, i.e., $\Lambda_{t_1,t_0}[\eins]=\eins$. Let us choose
in addition an Ohmic spectrum, $J(\omega)=\lambda \omega \exp(\omega/\omega_c)$ with the cutoff $\omega_c\gg\wS$. After some algebra and a transformation back to the Schr\"odinger picture, \eqnref{eq:TCLME} yields
\begin{eqnarray}
\dd{t}\rho_S(t) = -i\left[\frac{\wS}{2}\sz,\rho_S(t)\right] + \gamma(t) \left(\sbar \rho_S(t) \sbar^\dagger -\rho_S(t) \right),
\label{eq:exME}
\end{eqnarray}
where the time-dependent noise rate is given by
\begin{equation}
\gamma(t)=\frac{\lambda}{\beta}\arctan(\omega_c t),
\end{equation}
and the resulting Lindblad operator in the dissipative part of the evolution is a linear combination of Pauli-operators,
\begin{equation}
\sbar = \cos\Cang \sx + \sin \Cang \sz.
\end{equation}
{By sending the cutoff frequency $\omega_c$ to infinity, we obtain
\begin{eqnarray}
\gamma \equiv  \lim_{\omega_c \rightarrow \infty} \gamma(t) = \frac{\lambda \pi}{2 \beta},
\label{eq:SemigroupLimit}
\end{eqnarray}
so that} in this limit the master equation is a proper Lindblad master equation and the resulting dynamics can hence be described by a CPTP dynamical map which obeys the semigroup composition law \eqref{eq:mapSemigroup}, and then it is time-homogeneous Markovian.
{Indeed, for any value $\Cang\neq k \pi/2$ with odd $k$ the master equation in \eqref{eq:exME}
gives a NPC evolution:
it includes non-secular
contributions, which would be removed by the secular approximation
leading instead to a master equation in the form as \eqnref{eq:SecMasterEquation};
finally, for $\Cang = k \pi$ with even $k$ one has
the special case of a
purely transversal noise (i.e. only one Lindblad operator, with direction orthogonal to that of the free
Hamiltonian) \cite{Chaves2013}.}

\subsection{Markovian {non-Monotonic Population and Coherence Dynamics}}

\begin{figure}[t!]
 \includegraphics[width=1\columnwidth]{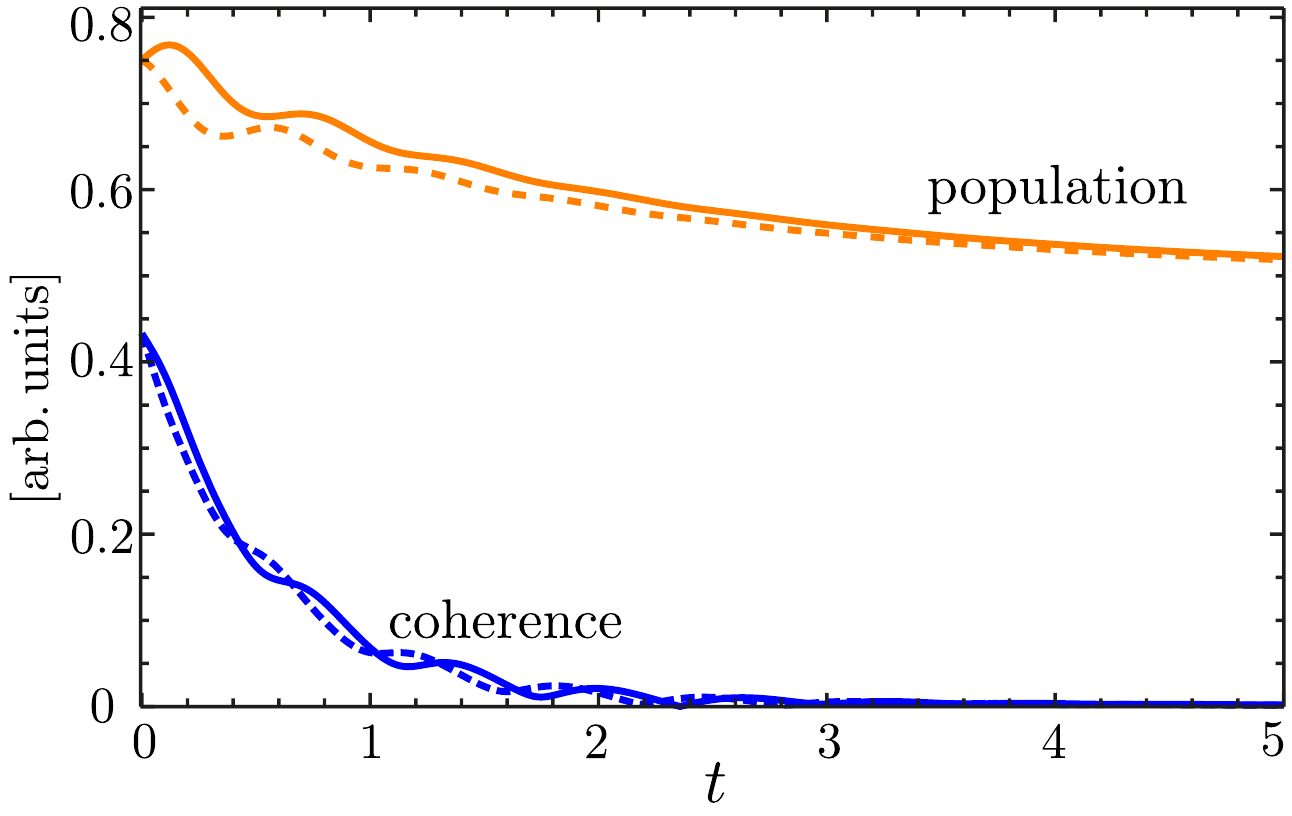}
\caption{%
Qubit population, $\langle 1|\rho_S(t) |1\rangle$, (orange) and coherence, $|\langle 1|\rho_S(t) |0\rangle|$, (blue)
for the initial phases $\varphi=0$ (solid) and $\varphi=\pi/2$ (dashed). As described in the main text, these clearly show the non-monotonic behavior and phase dependence of NPC semigroup dynamics. Parameters for the simulation: $\wS =10,\,\gamma = 1$, $\Cang = \theta = \pi/3$.}
\label{fig:evo}
\end{figure}

Let us investigate the dynamics induced by the master equation \eqref{eq:exME},
{in the semigroup limit fixed by \eqnref{eq:SemigroupLimit}}, for initially pure states {parametrized as}
\begin{eqnarray}
\rho(t_0) = \frac{\eins + \sin\theta(\cos\varphi\sx+\sin\varphi\sy) + \cos\theta\sz}{2}.
\label{eq:IniStateAngle}
\end{eqnarray}
For this purpose, we choose $\Cang = \theta = \pi/3$.
It is instructive to illustrate
{what happens for two different initial states, with} phases $\varphi_1=0$ and $\varphi_2=\pi/2$, respectively;
{the two corresponding evolutions
of the population $p(t)=\langle 1|\rho_S(t) |1\rangle$ and the absolute value of the
off-diagonal element $c(t)=\langle 1|\rho_S(t) |0\rangle$ are shown in \figref{fig:evo}; note that in the following we will often refer
to $|c(t)|$ directly as the coherence, since it is the quantity whose monotonicity we are
interested in and it can also be seen as a proper quantifier of the amount of coherence \cite{Baumgratz2014}
in the state of a two-level system.}

Neither the population nor the coherence show a monotonic {behavior but, instead,
we can observe}
damped oscillating evolutions, which are even different for the two chosen initial phases.
{We emphasize that such oscillations appear despite being
in the presence of a semigroup, i.e., a Markovian
time-homogeneous, dynamics. For example, this implies that the evolution of the
trace distance $d(t)$ defined as in \eqnref{eq:td} is monotonically non-increasing
for any couple of initial states, so that, in particular, the non-monotonicity appearing in all the population
and coherence evolutions for the two states reported in \figref{fig:evo} would
cancel out when evaluating their trace distance.}

{We note that the different qualitative features in the
evolutions shown in \figref{fig:evo} for the different phases of the initial states
are indeed a consequence}
of the non isotropic contraction of the Bloch sphere induced by the NPC dynamics, i.e.,
by the non-commu\-ta\-tivity of the free evolution and the action of the noise.
{More in general, the latter
implies that the effects of the free and dissipative parts of the evolution cannot be separated unambiguously.
Of course, this hinders a clear intuition about the different features of the dynamics
and, especially, which of them can be assigned to memory effects
due to the interaction with the environment and which of them are instead more
related to the leakage of information out of the open system.
As will become apparent in the following analysis, such an intuition can be recovered in the
regime of PC noise}.

\subsection{Monotonicity in Markovian Phase-Covariant Dynamics}
\begin{figure}[t!]
\begin{center}
\includegraphics[width=0.85\columnwidth]{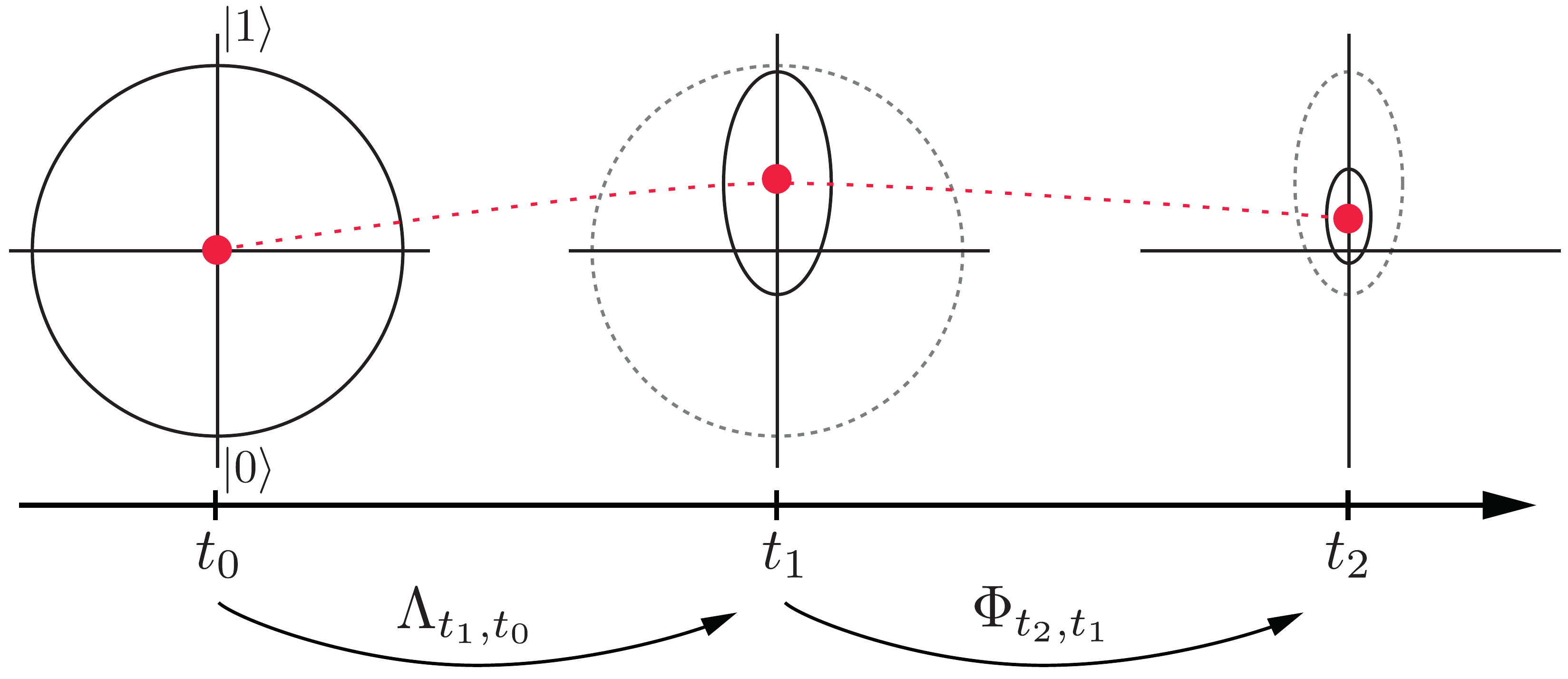}
\caption{%
Example of a CP-divisible, PC dynamics showing a non-monotonic evolution of the population:
$z-y$ section of the Bloch sphere at three different instants of time $t_0< t_1< t_2$
(as a consequence of PC, the whole ellipsoid at any time is simply obtained by rotating
the $z-y$ section about the $z$-axis);
the red dot denotes the state evolved from $\mathbbm{1}/2$ and the red dashed line the corresponding
$z$ component, $S_z(t)$.
CP divisibility, i.e. the CP of the propagators connecting the Bloch sphere at the different times,
implies that at every time the image of the Bloch sphere has to be
contained in the image at a previous time (which is reported in the figure with dotted lines),
and also some stronger conditions given in \cite{King2001};
for our purposes, it is enough to say that some constraints are set on the axes length
and center position of the ellipsoids at different times,
and they are satisfied in the figure.}
\label{fig:PCpositivity}
\end{center}
\end{figure}
{As mentioned in the Introduction,}
Markovianity is often associated with a monotonic behaviour of populations and coherences.
{As we have shown in the previous paragraph, this
cannot be justified in general, even in the case of a semigroup evolution;
however, we will now argue that such an intuition can be recovered, by adding some further constraint on the dynamics.}

First, recall that Markovianity is defined via the composition law \eqref{eq:mapComposition}, where the propagator
$\Phi_{t_2,t_1}$ is always CP.
{As said, this property is equivalent (under some regularity conditions
on the map) to}
the positivity of the rates in the master equation \eqref{eq:ME}.
{Moreover, it is easy to see from \eqnref{eq:PC} that the evolution of the coherence
under a PC evolution is simply given by
\begin{equation}\label{eq:ct}
|c(t)| = e^{\Gamma(t)} |c(0)|=
e^{ -\frac{1}{2} \int_{t_0}^t \left[\gamma_+(\tau) + \gamma_-(\tau) +4 \gamma_z(\tau)\right] \,\mathrm{d}\tau } |c(0)|,
\end{equation}
where we used explicitly \eqnref{eq:generalSolsPC}.
It is then clear that any PC, Markovian dynamics will be characterized by a monotonic decay of the coherence.
In addition, if we restrict to the case of a semigroup dynamics, i.e., we have positive time-independent
$\gamma_i(t)=\gamma_i \geq 0$, for $i=\pm,z$, the decay is even exponential, with rate
\begin{equation}
\gamma =\frac{\gamma_++\gamma_-+4\gamma_z}{2}.
\end{equation}

The situation is a bit more complicated for what concerns the evolution of the population.
For any PC dynamics, we have from \eqnref{eq:PC} (see also \cite{Teittinen2018}) that
the $z$ component of the Bloch vector is mapped according to}
\begin{equation}\label{eq:szt}
S_z(t_0)\mapsto S_z(t) = \kappa(t) + e^{\delta(t)} S_z(t_0);
\end{equation}
of course, the population $p(t)=(1+S_z(t))/2$ will have the same kind of (non-)monotonic behavior
as $S_z(t)$. Now, in general, depending on the choice of the initial condition $S_z(t_0)$, the quantity in \eqnref{eq:szt}
might exhibit a non-monotonic behavior even for positive rates $\gamma_i(t)$.
An example is given in Fig.\ref{fig:PCpositivity}, where we report the evolution of the Bloch sphere
at three different instants of time of a Markovian dynamics.
It is clear that CP divisibility implies that at every time the image of the Bloch sphere
must be contained in the image at the previous times (i.e. the propagators have to be positive).
In particular this means that the pure states initially at the antipodal points along the $z$-axis
have to decrease monotonically the length of their component along the $z$-axis itself
(i.e. the population will be monotonically non-decreasing or non-increasing for the initial state $\ket{0}$
and $\ket{1}$, respectively). However, if we take a mixture of the two, e.g., the fully mixed state $\mathbbm{1}/2$,
the corresponding Bloch vector might still evolve first toward one direction and after toward the opposite one,
as shown in Fig.\ref{fig:PCpositivity},
which implies of course a non-monotonic evolution of the population.
Clearly, such a behavior relies on the inhomogeneity of the dynamics, i.e., on the fact that the propagators
connecting states at different points in time are different (even if the length
of the time intervals is the same). In fact, one can easily see that in the case of a PC Markovian
time-homogeneous (i.e., semigroup) dynamics,
also the evolution of the population has to be monotonic, since in this case
\eqnref{eq:szt} reduces to (see \eqnref{eq:generalSolsPC})
\begin{equation}\label{eq:szt2}
S_z(t_0)\mapsto S_z(t) = \frac{\gamma_+-\gamma_-}{\gamma_+ + \gamma_-}(1-e^{-(\gamma_++\gamma_-)(t-t_0)})
+e^{-(\gamma_++\gamma_-)(t-t_0)}S_z(t_0),
\end{equation}
whose monotonicity is fixed once and for all by the
initial condition: we have a monotonic
decreasing (increasing) population iff $S_z(t_0)>\frac{\gamma_+ - \gamma_-}{\gamma_+ + \gamma_-}$
($S_z(t_0)<\frac{\gamma_+-\gamma_-}{\gamma_++\gamma_-}$).

Summarizing, we have shown that a qubit Markovian PC dynamics is always characterized by a monotonic decay
of the coherence, while the monotonic behavior of the population is guaranteed
only by the stronger requirement of a Markovian time-homogeneous dynamics.
As a consequence, we can also conclude that the non-monotonic behavior of
both the population and the coherence
described in the previous paragraph is due to the NPC nature
of the semigroup dynamics considered there.

\subsection{Role of multiple Initial Conditions}
\label{sec:PCequivalence}

\begin{figure}[t!]
 \includegraphics[width=1\columnwidth]{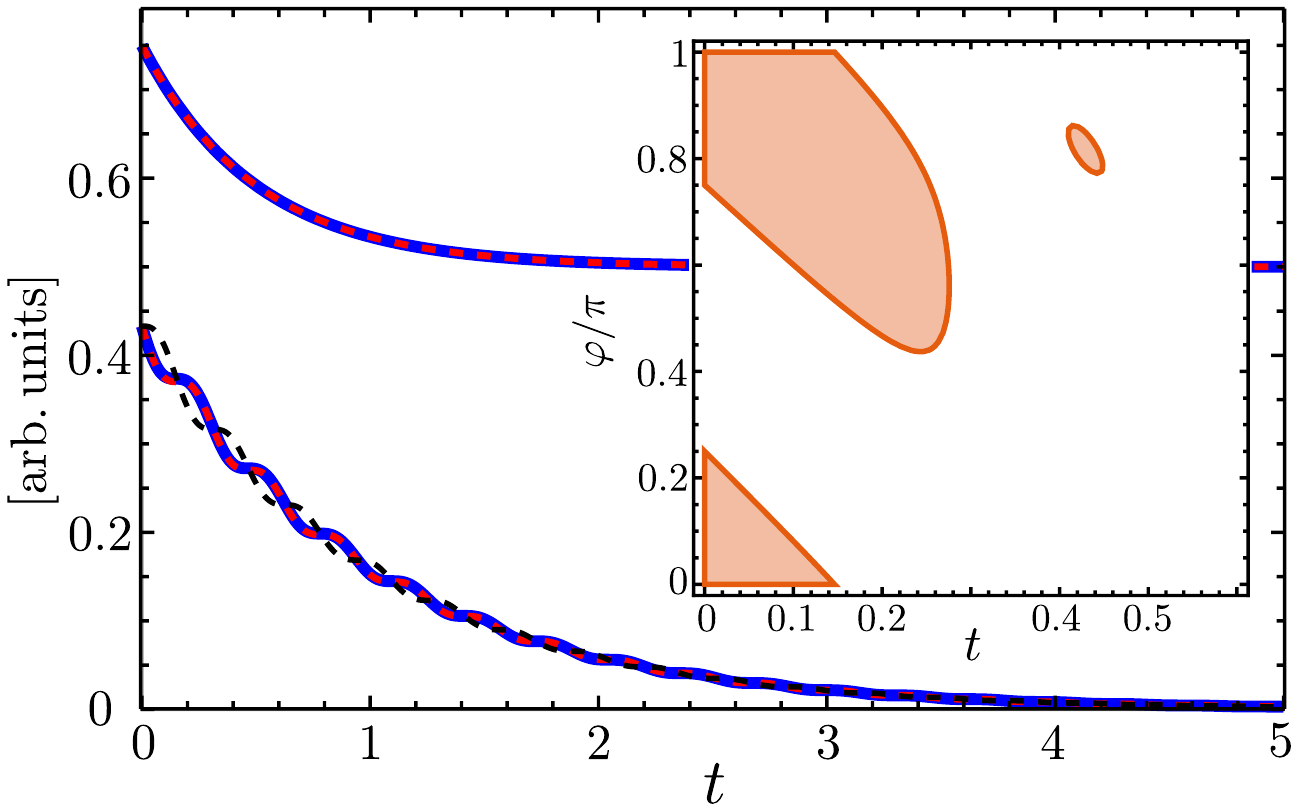}
\caption{%
Artificial PC evolution mimicking an NPC dynamics. The thick blue lines show $p(t)$ and $|c(t)|$
for the NPC dynamics,
see \eqnref{eq:NPCsol}, for $\varphi=\pi/2$ and $\theta=\pi/3$ with $\wS=10$ and $\gamma=1$. The red dashed lines
show the population and coherence for the artificially created PC dynamics as in Eqs.(\ref{eq:PCsol})
and (\ref{eq:fix}). The black dashed lines illustrates the NPC $|c(t)|$ for the initial state
with phase $\varphi'=\pi$;
note that the artificial PC map fixed by $\varphi=\pi/2$ would not be sensitive to
a change of phase in the initial state it is applied to. The inset shows the regions where one of eigenvalues of the Choi state
of the artificial PC map is smaller than zero.
Parameters for the simulation: $\wS =10,\,\gamma = 1, \Cang=0$.}
\label{fig:artificialDynamics}
\end{figure}

Non-monotonic evolutions of both populations and coherences indeed occur
also in PC dynamics, when some memory effects break the CP-divisibility.
Here, we want to show how some non-monotonic behaviors of the coherence
obtained in a semigroup NPC dynamics can be reproduced via PC
evolutions, if one introduces memory effects which depend on the specific choice of the initial state.

In order to keep all the calculations analytical, we consider the master equation \eqref{eq:ME}
in the semigroup limit, see \eqnref{eq:SemigroupLimit},
in the special case of the transversal noise $\Cang = 0$.
The resulting evolution is given by
\begin{eqnarray}
p(t) &=&\frac{1}{2}(1+\cos \theta e^{-2 \gamma t})\notag \\
c(t) &=& \frac{\sin(\theta)}{2} \,e^{-\gamma t-i \varphi} \left(\cosh\left( t\sqrt{\gamma^2-\wS^2}\right) +\frac{e^{2i\varphi}\gamma - i\wS}{\sqrt{\gamma^2-\wS^2}} \sinh\left( t\sqrt{\gamma^2-\wS^2}\right) \right).
\label{eq:NPCsol}
\end{eqnarray}
We thus have an exponential monotonic decay of the population, while, for $\omega_0 > \gamma$,
the coherence $|c(t)|$ undergoes damped oscillations.

In the following, we aim to simulate this dynamics via PC evolutions, extracting the necessary decay rates and the Lamb shift to plug into \eqnref{eq:SecMasterEquation}.
Now, since the NPC map is unital (translations of the Bloch sphere are suppressed by the high-temperature limit),
the same should be for the corresponding PC evolutions, so that we have to set $\kappa(t)=0$.
Exploiting this relation
and using the form of
the initial state in \eqref{eq:IniStateAngle}, the general solution
for a PC evolution is
\begin{eqnarray}
p_{\mathrm{PC}}(t) &=& \frac{1}{2}(1+\cos \theta e^{\delta(t)})\notag \\
c_{\mathrm{PC}}(t) &=& \frac{\sin(\theta)}{2}\, e^{\Gamma(t)-i \phi(t)}.
\label{eq:PCsol}
\end{eqnarray}
Comparing the population dynamics in \eqnref{eq:NPCsol} and \eqnref{eq:PCsol}, we immediately find $\gamma_+(t) = \gamma_-(t) = \gamma$. Now, separating the contributions
from the absolute value and the phase in $c(t)$
and considering $\gamma\ll\wS$,
we find that the rates in the PC master equation \eqref{eq:SecMasterEquation}
reproducing the behavior in \eqnref{eq:NPCsol} to first order in $\gamma$
are given by
\begin{eqnarray}
\gamma_+(t) &=& \gamma_-(t) = \gamma, \notag \\
\gamma_z(t) &=& - \gamma \cos (2\wS t +2\varphi), \notag \\
h(t) &=& - \gamma \sin (2 \wS t+2\varphi).\label{eq:fix}
\end{eqnarray}
The solution of this artificial PC model, as well as the NPC semigroup dynamics are illustrated in \figref{fig:artificialDynamics},
where we can observe the match between the two.

Let us now comment on a few peculiarities of the PC description.
First of all, this model is highly non-Markovian, since the rate $\gamma_z(t)$ is obviously negative for multiple intervals of the evolution time, so that, as expected, in order to reproduce the coherence oscillations
appearing in the semigroup NPC dynamics with a PC evolution we need to introduce memory effects.
Moreover, one should note the dependence of the PC rates on the phase $\varphi$ of the initial state.
The latter is necessary to emulate the
dependence of the coherence evolution on the initial phase, typical of NPC dynamics.
However, this dependence of course introduces a non-linearity on the effective PC maps
we are using to describe the coherence evolution.
This means that such description cannot be inferred from a microscopic
model of the system-environment interaction, fixed by a linear Hamiltonian,
but should rather seen as a phenomenological
characterization based on the "ad-hoc" introduction of the parameter $\varphi$.
Most importantly,
it is rather clear that the map given by \eqnref{eq:fix}
for a certain value of $\varphi$
would not describe properly the evolution of the coherence if
we apply it to an initial state with a different phase, $\varphi'$,
as exemplified by the black line in \figref{fig:artificialDynamics}.
This clearly shows that both PC and Markovianity of the dynamics are properties
which cannot be assessed by monitoring one single evolution
of the open-system state for a fixed initial condition. Instead, one needs the comparison
between the evolutions from at least two different initial conditions to have the possibility
to witness the violation of such properties.
The dependence of the evolution on the phase of the initial state witnesses
the NPC of the dynamics, while, for example, the non-monotonicity of the trace distance
witnesses its non-Markovianity.

Finally, let us briefly comment on the CP of the artificial PC description.
While the map defined by \eqnref{eq:fix} for a fixed $\varphi$
is always positive and hence represents a physical evolution,
there can exist time intervals where the dynamics is not CP.
These areas depend on the value on $\varphi$
and are reported in the inset of \figref{fig:artificialDynamics}.
Crucially, there exists a region of $\varphi$, where the map is indeed CP for all times, while for other values CP
is only violated at short times.

\section{Conclusions}
In this work we have provided a simple model which illustrates how the observation of non-monotonicity in the qubit's state properties, namely its population and coherence, is by no means a witness of non-Markovian dynamics.
Importantly, our model is microscopically motivated and it is even described by a
a semigroup of completely positive maps.
We further clarified that
the connection of population and coherence monotonicity
with Markovianity can be established
when the system is undergoing a PC dynamics;
in this case, Markovianity (CP-divisibility) implies
a motononic decay of coherences, while the stronger assumption of
time-homogeneous Markovianity (i.e. semigroup) guarantees
a monotonic behaviour also for populations.
In addition,
given the non-monotonic
decay of the coherence in a semigroup NPC dynamics,
we showed that it is possible to formulate a PC phenomenological model mimicking
such behavior, but at the price of
introducing memory effects which depend on the choice of the initial state.
This also illustrates the necessity to consider the state evolution for more than
one fixed initial condition when one wants to have a meaningful
witness of non-Markovianity.

We hope that this discussion shed some light onto the peculiarities of NPC evolutions
and further clarifies why it is not possible to provide any general (i.e., model-independent) definition of non-Markovianity
on the basis of the observed dynamics for a single initial preparation.

\paragraph{Acknowledgments:}
We would like to thank Bassano Vacchini, Heinz-Peter Breuer and Angelo Bassi for the organization of the 684. WE-Heraeus-Seminar on Advances in open systems and fundamental tests of quantum mechanics and very specially to the Wilhelm und Else Heraeus Stiftung for their continuous support of fundamental research in Physics.


\begin{thebibliography}{10}
\bibitem{Breuer2002}H.-P. Breuer and F. Petruccione, \emph{The Theory of Open Quantum
Systems}, (Oxford University Press, Oxford, 2002).
\bibitem{Rivas2012} {\'A}. Rivas and S. F. Huelga, \emph{Open Quantum Systems}, (Springer, New York, 2012).
\bibitem{Rivas2014}{\'A}. Rivas, S. F. Huelga, andM. B. Plenio, Rep. Progr.
Phys. {\bf77}, 094001 (2014),
\url{http://stacks.iop.org/0034-4885/77/i=9/a=094001}.
\bibitem{Breuer2016}H.-P. Breuer, E.-M. Laine, J. Piilo, and B. Vacchini, Rev.
Mod. Phys. {\bf88}, 021002 (2016),
\url{https://link.aps.org/doi/10.1103/RevModPhys.88.021002}.
\bibitem{deVega2017} I. de Vega and D. Alonso, Rev. Mod. Phys. {\bf89}, 015001 (2017),
\url{https://link.aps.org/doi/10.1103/RevModPhys.89.015001}.
\bibitem{Li2017}L. Li, M. J. W. Hall, and H. M. Wiseman, Phys. Rep. {\bf759}, 1 (2018),
\url{https://doi.org/10.1016/j.physrep.2018.07.001}.
\bibitem{Laine2014}
E.-M. Laine, H.-P. Breuer, and J. Piilo, Scientific Reports
{\bf4}, 4620 (2014), \url{http://dx.doi.org/10.1038/srep04620}.
\bibitem{Torre2018} G. Torre and F. Illuminati, ArXiv e-prints (2018), 1805.03617.
\bibitem{Vasile2011}R. Vasile, S. Olivares, M. A. Paris, and S. Maniscalco, Phys. Rev. A {\bf83}, 042321 (2011),
\url{https://link.aps.org/doi/10.1103/PhysRevA.83.042321}.
\bibitem{Liu2016}B.-H. Liu, X.-M. Hu, Y.-F. Huang, C.-F. Li, G.-C. Guo,
A. Karlsson, E.-M. Laine, S. Maniscalco, C. Macchiavello, and
J. Piilo, Eurphys. Lett. {\bf114}, 10005 (2016), \url{http://stacks.iop.org/0295-5075/114/i=1/a=10005}.
\bibitem{Bylicka2014}B. Bylicka, D. Chr{\'u}sc{\'i}nski, and S. Maniscalco, Scientific Reports
{\bf4}, 5720 (2014), \url{http://dx.doi.org/10.1038/srep05720}.
\bibitem{Plenio1999} M. B. Plenio, S. F. Huelga, A. Beige, and P. L. Knight, Phys.
Rev. A {\bf59}, 2468 (1999), \url{https://link.aps.org/doi/10.1103/PhysRevA.59.2468}.
%
\bibitem{Plenio2008} S. F. Huelga, A. Rivas, and M. B. Plenio
Phys. Rev. Lett. {\bf 108}, 160402 (2012), \url{https://doi.org/10.1103/PhysRevLett.108.160402}
%
\bibitem{Cywinski2008} L. Cyw{\'i}nski, R. M. Lutchyn, C. P. Nave, and S. Das Sarma,
Phys. Rev. B {\bf77}, 174509 (2008), \url{http://link.aps.org/doi/10.1103/PhysRevB.77.174509}.
\bibitem{Lange2011}G. de Lange, D. Rist{\'e}, V. V. Dobrovitski, and R. Hanson, Phys. Rev. Lett.
{\bf106}, 080802 (2011), \url{http://link.aps.org/doi/10.1103/PhysRevLett.106.080802}.
\bibitem{Degen2017}C. L. Degen, F. Reinhard, and P. Cappellaro, Rev. Mod. Phys.
{\bf89}, 035002 (2017), \url{https://link.aps.org/doi/10.1103/RevModPhys.89.035002}.
%
\bibitem{Chin2012}
A.W. Chin, S. F. Huelga, and M. B. Plenio, Phys. Rev. Lett. {\bf109}, 233601 (2012),  
\url{https://doi.org/10.1103/PhysRevLett.109.233601}.
\bibitem{Benjamin2011} Y. Matsuzaki, S. C. Benjamin,  and J. Fitzsimons, Phys. Rev. A {\bf 84}, 012103 (2011),
\url{https://doi.org/10.1103/PhysRevA.84.012103}
%
\bibitem{Smirne2016}A. Smirne, J. Ko{\l}od{\'y}nski, S. F. Huelga, and R. Demkowicz-Dobrza{\'n}ski, Phys. Rev. Lett. {\bf116}, 120801 (2016), \url{http://link.aps.org/doi/10.1103/PhysRevLett.116.120801}.
\bibitem{Haase2018Metro}J. F. Haase, A. Smirne, J. Ko{\l}od{\'y}nski, R. Demkowicz-Dobrza{\'n}ski, and S. F. Huelga, New J. Phys.
{\bf20}, 053009 (2018), \url{http://stacks.iop.org/1367-2630/20/i=5/a=053009}.
\bibitem{Liu2011}B.-H. Liu, L. Li, Y.-F. Huang, C.-F. Li, G.-C. Guo, E.-M. Laine,
H.-P. Breuer, and J. Piilo, Nature Physics {\bf7}, 931 (2011),
\url{http://dx.doi.org/10.1038/nphys2085}.
\bibitem{Cialdi2011}S. Cialdi, D. Brivio, E. Tesio, and M. G. A. Paris, Phys. Rev. A
{\bf83}, 042308 (2011), \url{https://link.aps.org/doi/10.1103/PhysRevA.83.042308}.
\bibitem{Chiuri2012}A. Chiuri, C. Greganti, L. Mazzola, M. Paternostro, and P. Mataloni,
Scientific Reports {\bf2}, 968 (2012) \url{https://doi.org/10.1038/srep00968}.
\bibitem{Liu2013}B.-H. Liu, D.-Y. Cao, Y.-F. Huang, C.-F. Li, G.-C. Guo, E.-M.
Laine, H.-P. Breuer, and J. Piilo, Scientific Reports {\bf3}, 1781
(2013), \url{http://dx.doi.org/10.1038/srep01781}.
\bibitem{Jin2015}J. Jin, V. Giovannetti, R. Fazio, F. Sciarrino, P. Mataloni,
A. Crespi, and R. Osellame, Phys. Rev. A {\bf91},
012122 (2015), \url{https://link.aps.org/doi/10.1103/PhysRevA.91.012122}.
\bibitem{Cialdi2017}S. Cialdi, M. A. C. Rossi, C. Benedetti, B. Vacchini, D. Tamascelli,
S. Olivares, and M. G. A. Paris, Appl. Phys. Lett. {\bf110}, 081107 (2017),
\url{https://doi.org/10.1063/1.4977023}.
\bibitem{Wittemer2018}M.Wittemer, G. Clos, H.-P. Breuer, U.Warring, and T. Schaetz,
Phys. Rev. A {\bf97}, 020102 (2018), \url{https://link.aps.org/doi/10.1103/PhysRevA.97.020102}.
\bibitem{Haase2018}J. F. Haase, P. J. Vetter, T. Unden, A. Smirne, J. Rosskopf,
B. Naydenov, A. Stacey, F. Jelezko, M. B. Plenio, and S. F.
Huelga, Phys. Rev. Lett. {\bf121}, 060401 (2018),
\url{https://link.aps.org/doi/10.1103/PhysRevLett.121.060401}.
\bibitem{Wang2018}F. Wang, P.-Y. Hou, Y.-Y. Huang, W.-G. Zhang, X.-L.
Ouyang, X. Wang, X.-Z. Huang, H.-L. Zhang, L. He, X.-Y.
Chang, and L.-M. Duan, Phys. Rev. B {\bf98}, 064306 (2018),
\url{https://doi.org/10.1103/PhysRevB.98.064306}.
\bibitem{Peng2018}
S. Peng, X. Xu, K. Xu, P. Huang, P. Wang, X. Kong, X.
Rong, F. Shi, C. Duan, and J. Du, Sci. Bull. {\bf63}, 336 (2018),
\url{https://doi.org/10.1016/j.scib.2018.02.017}
\bibitem{Tanimura1989}
Y. Tanimura and R. Kubo, J. Phys. Soc. Jap. {\bf58}, 1199 (1989),
\url{https://journals.jps.jp/doi/10.1143/JPSJ.58.1199};
Z. Tang, X. Ouyang, Z. Gong, H. Wang, and J. Wu, J. Chem. Phys. {\bf143}, 224112 (2015),
\url{https://doi.org/10.1063/1.4936924}.
\bibitem{Diosi1998}
L. Diosi, W. T. Strunz, and N. Gisin, Phys. Rev. A {\bf58}, 1699 (1998),
\url{https://doi.org/10.1103/PhysRevA.58.1699}.
\bibitem{Prior2010}
J. Prior, A. W. Chin, S. F. Huelga, and M. B. Plenio, Phys. Rev. Lett. {\bf105}, 050404 (2010)
\url{https://doi.org/10.1103/PhysRevLett.105.050404}.
\bibitem{Ciccarello2013}
F. Ciccarello, G. M. Palma, and V. Giovannetti, Phys. Rev. A {\bf87}, 040103(R) (2013),
\url{https://doi.org/10.1103/PhysRevA.87.040103}.
\bibitem{Diosi2014}
L. Diosi and L. Ferialdi, Phys. Rev. Lett. {\bf113}, 200403 (2014),
\url{https://doi.org/10.1103/PhysRevLett.113.200403};
L. Ferialdi, Phys. Rev. Lett. {\bf116}, 120402 (2016),
\url{https://doi.org/10.1103/PhysRevLett.116.120402};
\bibitem{Batalhao2014}
T. B. Batalhao,G. D. de Moraes Neto, M. A. de Ponte, and M. H. Y. Moussa,
Phys. Rev. A {\bf90}, 032105 (2014),
\url{https://doi.org/10.1103/PhysRevA.90.032105}.
\bibitem{Vacchini2016}
B. Vacchini, Phys. Rev. Lett. {\bf117}, 230401 (2016),
\url{https://doi.org/10.1103/PhysRevLett.117.230401}.
\bibitem{Gasbarri2017}
G. Gasbarri, M. Toro\v{s}, and A. Bassi, Phys. Rev. Lett. {\bf119} 100403 (2017),
\url{https://doi.org/10.1103/PhysRevLett.119.100403};
G. Gasbarri and L. Ferialdi, Phys. Rev. A {\bf98}, 042111 (2018)
\url{https://doi.org/10.1103/PhysRevA.98.042111}.
\bibitem{Tamascelli2018}
D. Tamascelli, A. Smirne, S. F. Huelga, and M. B. Plenio,
Phys. Rev. Lett. {\bf120}, 030402 (2018)
\url{https://doi.org/10.1103/PhysRevLett.120.030402};
D. Tamascelli, A. Smirne, S. F. Huelga, and M. B. Plenio, arXiv:1811.12418.
\bibitem{Campbell2018}
S. Campbell, F. Ciccarello, G. M. Palma and B. Vacchini, Phys. Rev. A {\bf98}, 012142 (2018),
\url{https://doi.org/10.1103/PhysRevA.98.012142}.
\bibitem{vanKampen1974}
N. G. van Kampen, Physica {\bf74}, 215 (1974)
\url{https://doi.org/10.1016/0031-8914(74)90121-9};
\emph{ibid.} 74, 239 (1974)
\url{https://doi.org/10.1016/0031-8914(74)90122-0}.
\bibitem{Shibata1977} F. Shibata, Y. Takahashi and N. Hashitume, J. Stat. Phys. {\bf17}, 171(1977)
\url{https://doi.org/10.1007/BF01040100}
\bibitem{Chaturvedi1979}S. Chaturvedi and F. Shibata, Z. Phys. B {\bf35}, 297 (1979)
\url{https://doi.org/10.1007/BF01319852"}.
\bibitem{Breuer2001}H.-P. Breuer, B. Kappler, and F. Petruccione, Ann. Phys. {\bf291}, 36 (2001),
\url{http://www.sciencedirect.com/science/article/pii/S0003491601961524}.
\bibitem{Rivas2010}{\'A}. Rivas, S. F. Huelga, and M. B. Plenio, Phys. Rev. Lett.
{\bf105}, 050403 (2010), \url{http://link.aps.org/doi/10.1103/PhysRevLett.105.050403}.
\bibitem{Breuer2009}H.-P. Breuer, E.-M. Laine, and J. Piilo, Phys. Rev. Lett.
{\bf103}, 210401 (2009), \url{http://link.aps.org/doi/10.1103/PhysRevLett.103.210401}.
\bibitem{Madsen2011}K. H. Madsen, S. Ates, T. Lund-Hansen, A. L{\"o}er, S. Reitzenstein,
A. Forchel, and P. Lodahl, Phys. Rev. Lett.
{\bf106}, 233601 (2011), \url{https://link.aps.org/doi/10.1103/PhysRevLett.106.233601}.
\bibitem{Duan2017}H.-G. Duan, V. I. Prokhorenko, R. J. Cogdell, K. Ashraf, A. L.
Stevens, M. Thorwart, and R. J. D. Miller, Proceedings of the
National Academy of Sciences 114, 8493 (2017), \url{https://www.pnas.org/content/114/32/8493}.
\bibitem{Passos2018}M. H. M. Passos, P. C. Obando, W. F. Balthazar, F. M.
Paula, J. A. O. Huguenin, and M. S. Sarandy, arXiv e-prints
arXiv:1807.05378 (2018).
\bibitem{Holevo1993}A. Holevo, Rep. Math. Phys. {\bf32}, 211 (1993),
\url{http://www.sciencedirect.com/science/article/pii/0034487793900146}.
\bibitem{Holevo1996}A. Holevo, J. Math. Phys. {\bf37}, 1812
(1996), \url{http://dx.doi.org/10.1063/1.531481}.
\bibitem{Vacchini2010}B. Vacchini, Lect. Notes Phys. {\bf787}, 39 (2010).
\bibitem{Gorini1976} V. Gorini, A. Kossakowski, and E. C. G. Sudarshan, J. Math. Phys. (N.Y.) {\bf17}, 821 (1976),
\url{https://doi.org/10.1063/1.522979}.
\bibitem{Hall2014}
M.J. W. Hall, J.D. Cresser, L. Li, and E. Andersson, Phys. Rev. A {\bf89}, 042120 (2014),
\url{https://doi.org/10.1103/PhysRevA.89.042120}.
\bibitem{Lindblad1976}
G. Lindblad, Commun. Math. Phys. {\bf48}, 119 (1976),
\url{https://doi.org/10.1007/BF01608499}.
\bibitem{Laine2010}
E.-M. Laine, J. Piilo, and H.-P. Breuer, Phys. Rev. A {\bf81}, 062115 (2010)
\url{https://doi.org/10.1103/PhysRevA.81.062115}.
\bibitem{Chruscinski2018}
D. Chru{\'s}ci{\'n}ski, {\'A}. Rivas, and E. St{\o}rmer,
Phys. Rev. Lett. {\bf121}, 080407 (2018)
\url{https://doi.org/10.1103/PhysRevLett.121.080407}
\bibitem{King2001}
C. King and M. Ruskai, IEEE Trans. Inf. Theory {\bf47}, 192 (2001),
\url{https://doi.org/10.1109/18.904522};
M. B. Ruskai, S. Szarek, and E. Werner, Linear Algebra Appl. {\bf347}, 159 (2002),
\url{https://doi.org/10.1016/S0024-3795(01)00547-X}.
\bibitem{Andersson2007}
E. Andersson, J. D. Cresser, and M. J.W. Hall, J. Mod. Opt. {\bf54}, 1695 (2007),
\url{https://doi.org/10.1080/09500340701352581}.
\bibitem{Smirne2010}
A. Smirne and B. Vacchini, Phys. Rev. A {\bf82}, 022110 (2010),
\url{https://doi.org/10.1103/PhysRevA.82.022110}.
\bibitem{Marvian2016}
I. Marvian and R.W. Spekkens, Phys. Rev. A {\bf94}, 052324 (2016),
\url{https://doi.org/10.1103/PhysRevA.94.052324}
\bibitem{Lostaglio2017}
M.Lostaglio, K. Korzekwa, and A. Milne, Phys. Rev. A {\bf96}, 032109 (2017),
\url{https://doi.org/10.1103/PhysRevA.96.032109}.
\bibitem{Leggett1987}A. J. Leggett, S. Chakravarty, A. T. Dorsey, M. P. A. Fisher,
A. Garg, and W. Zwerger, Rev. Mod. Phys. {\bf59}, 1 (1987),
\url{http://link.aps.org/doi/10.1103/RevModPhys.59.1}.
\bibitem{Chaves2013}
R. Chaves, J.B. Brask, M. Markiewicz, J. Ko{\l}ody{\'n}ski and A. Ac{\'i}n, Phys. Rev. Lett. {\bf111} 120401 (2013),
\url{https://doi.org/10.1103/PhysRevLett.111.120401};
J.B. Brask, R. Chaves, and J. Ko{\l}ody{\'n}ski, Phys. Rev. X {\bf 5}, 031010 (2015)
\url{https://doi.org/10.1103/PhysRevX.5.031010}.
\bibitem{Baumgratz2014}
T. Baumgratz, M. Cramer, M. B. Plenio, Phys. Rev. Lett. {\bf113}, 140401 (2014)
\url{https://doi.org/10.1103/PhysRevLett.113.140401}.
\bibitem{Teittinen2018} J. Teittinen, H. Lyyra, B. Sokolov and S Maniscalco,
New J. Phys. {\bf20} 073012 (2018)
\url{https://doi.org/10.1088/1367-2630/aacc38}.
\end{thebibliography}
\end{document}